\documentclass[twocolumn,showpacs,preprintnumbers,amsmath,amssymb]{revtex4}

\usepackage{graphicx}
\usepackage{dcolumn}
\usepackage{bm}

\begin{document}

\title{Generation of GHZ-type and \emph{W}-type entangled coherent states of three-cavity fields}
\author{Chun-Hua Yuan}
\email{chunhuay@sjtu.edu.cn}
\author{Yong-Cheng Ou}
\author{Zhi-Ming Zhang}
\email{zzm@sjtu.edu.cn}
\affiliation{%
Department of Physics, Shanghai Jiao Tong University, Shanghai 200240, People's Republic of China}%
\date{\today}
\begin{abstract}
We present experimental schemes to prepare the three-cavity
GHZ-type and \emph{W}-type entangled coherent states in the
context of dispersive cavity quantum electrodynamics. The schemes
can be easily generalized to prepare the GHZ-type and
\emph{W}-type entangled coherent states of $n$-cavity fields. The
discussion of our schemes indicates that it can be realized by
current technologies.
\end{abstract}
\maketitle
Quantum entanglement is at the heart of quantum information
science and technology. Information may be coded on quantum
two-level systems ("qubits") \cite{Divincenzo}. Nonlocal
correlations between two qubits can be used for quantum key
distribution \cite{Ekert} or quantum teleportation
\cite{Bennett93}. More complex entanglement manipulations could be
used for quantum error correction \cite{Steane} or entanglement
purification \cite{VanEnk}. The preparation of complex quantum
entangled states in well-controlled conditions is the subject of
an intense experimental activities. Complex entangled states, such
as Greenberger-Horne-Zeilinger (GHZ) \cite{GHZ} class and the
\emph{W} \cite{W} class, can be used to understand basic quantum
phenomena and can also be used for information processing.
Recently, many schemes have been proposed for generating GHZ
states \cite{Bouwmeester}, and \emph{W} states \cite{Nikolai}.

Entanglement in continuous variables has been of great interest
since the celebrated paper of Einstein, Podolsky and Rosen (EPR)
\cite{Einstein} who constructed a two-particle state which was
strongly entangled both in position and momentum spaces. The
so-called continuous-variable quantum information is that quantum
information can be coded in a state which is characterized by
infinite number of degrees of freedom, such as a position or
momentum wave function of a microscopic particle or the quadrature
components of a field. The continuous-variable approach promises
to be more compact and more efficient in both coding and
manipulating quantum information and thus has been developed
rapidly during the last few years from both theoretical and
experimental point of view \cite{Vaidman}. An intermediate, quite
simple but very useful, way  for coding is to utilize
superpositions of a finite number of macroscopically
distinguishable states each of which is however embedded in an
unbounded vector space \cite{Cochrane}. In this approach, instead
of qubits, one deals with logical qubits. A logical qubit is
regarded as a superposition of two continuous-variable states
which are linearly independent but not necessarily orthogonal to
each other. An elegant choice for representing logical qubit is to
use two coherent states $|\alpha\rangle$ and $|-\alpha\rangle$ of
an optical mode, with $\alpha$ the complex coherent amplitude. A
coherent field is a fundamental tool in quantum optics and linear
superposition of two coherent states is considered one of the
realizable mesoscopic quantum systems \cite{Yurke}. To process
quantum information encoded in logical qubits, recently, Nguyen
\cite{Nguyen} firstly proposed GHZ-type and \emph{W}-type
entangled coherent states (ECSs). The so-called GHZ-type ECS is
\begin{eqnarray}
\label{e1} |\texttt{GHZ},\alpha\rangle_{1\ldots
N}=c_1|\alpha,\alpha,\ldots
,\alpha\rangle_{1\ldots N}\nonumber\\
+c_2|-\alpha,-\alpha,\ldots ,-\alpha\rangle_{1\ldots N},
\end{eqnarray}
where $c_{1,2}$ the normalization coefficients, and the
\emph{W}-type ECS is
\begin{eqnarray}
\label{e2} |\emph{W},\alpha\rangle_{1\ldots
N}=b_1|\alpha,-\alpha,\ldots
,-\alpha\rangle_{1\ldots N}\nonumber\\
+b_2|-\alpha,\alpha,\ldots ,-\alpha\rangle_{1\ldots N}+\ldots\nonumber\\
+b_N|-\alpha,-\alpha,\ldots ,\alpha\rangle_{1\ldots N},
\end{eqnarray}
where $b_i$ $(i=1,2,\ldots,N)$ the normalization coefficients
\cite{Nguyen}. Since the amount of entanglement in Eq.~(\ref{e1})
and Eq.~(\ref{e2}) is independent of $|\alpha\rangle$, it may seem
that such states may be especially robust against decoherence due
to photon absorption. Namely, although photon absorption
attenuates the coherent state, that by itself has no effect on the
entanglement.

In this paper, we propose schemes to generate the GHZ-type and
\emph{W}-type ECSs of three-cavity fields. Some theoretical
schemes have been proposed \cite{Gerry,ChuiYang} to generate
GHZ-type ECS, especially, Gerry \cite{Gerry} proposed an elegant
scheme to generate GHZ-type ECS of three cavities using a
$\Xi$-type three-level atom with three separated cavities, but
there is no report of experimental realization of the GHZ-type ECS
state. Here, we present an alternative, and feasible scheme to
generate GHZ-type ECS with current cavity QED technology. To our
best knowledge, our scheme here is the first one for generating
\emph{W}-type ECS. Our schemes are based on the present cavity
quantum electrodynamics (QED) techniques. Cavity QED, with Rydberg
atoms crossing superconducting cavities, offers an almost ideal
system for the generation of entangled states and implementation
of small scale quantum information processing.

We consider a system composed of a single two-level atom and a
single-mode cavity field. The atom-field interaction is described
by the Jaynes-Cummings model. In the rotating-wave approximation,
the Hamiltonian of the system is given by (assuming $\hbar=1$)
\cite{Scully}
\begin{eqnarray}
H=H_0+H_I,\\
H_0=\omega a^{\dag}a+\frac{1}{2}\omega_0\sigma_Z,\\
H_I=g(a^{\dag}\sigma^{-}+a\sigma^{+}),
\end{eqnarray}
where $\sigma^{+}=|e\rangle\langle g|$,
$\sigma^{-}=|g\rangle\langle e|$, and $\sigma_{z}=|e\rangle\langle
e|-|g\rangle\langle g|$. $a^{\dag}$ ($a$) is the creation
(annihilation) operator of the cavity field. $\omega_0$ and
$\omega$ are the atomic transition frequency and the cavity field
frequency, respectively, and $g$ is the atom-field coupling
constant.

\begin{figure}[hbtp!]
\centerline{\includegraphics[scale=0.8,angle=0]{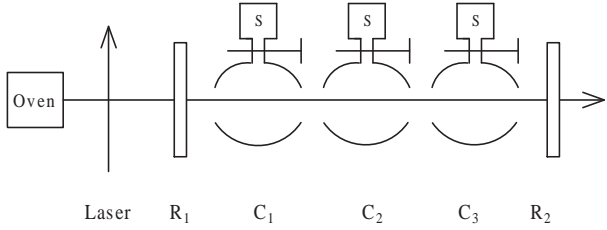}}
\caption{Experimental setup for generation of three-cavity
GHZ-type entangled coherent state. $C_1$, $C_2$, and $C_3$ are
identical cavities. $R_1$ and $R_2$ are the Ramsey zones. $S$ is a
classical source of microwaves which are injected into the cavity
through a waveguide.}\label{fig1}
\end{figure}

If the detuning between the atomic transition frequency and the
cavity field frequency is much larger than the coupling constant,
the atom has a negligible probability of making a transition
between the ground and excited states. Under this condition, the
effective interaction Hamiltonian can be written as \cite{Holland}
\begin{equation}
H_{eff}=\lambda a^{\dag}a\sigma_z,
\end{equation}
where $\lambda=\frac{g^2}{\delta}$, $\delta=\omega_0-\omega$. In
the interaction picture, the evolution of the whole system is
$|\psi(t)\rangle=e^{-iH_{eff}t}|\psi(0)\rangle$ ($[H_{eff},
H_0]=0$), where $|\psi(0)\rangle$ is the initial state of the
whole system. If the atom is initially in the excited state or in
the ground state and the cavity field is initially in the coherent
state $|\alpha\rangle$, then after the interaction time $\tau$ ,
the process from the initial state of the atom-field system to the
final state can be written as
\begin{equation}
|e\rangle|\alpha\rangle\rightarrow|e\rangle|\alpha
e^{-i\lambda\tau}\rangle,
\end{equation}
or
\begin{equation}
|g\rangle|\alpha\rangle\rightarrow|g\rangle|\alpha
e^{i\lambda\tau}\rangle.
\end{equation}

Firstly, we describe the process of preparing the GHZ-type
three-cavity ECS. The experimental setup for the proposed method
is shown in Fig.~\ref{fig1}. The three identical cavities $C_1$,
$C_2$, $C_3$ are prepared in the coherent state
$|\alpha\rangle_1$, $|\alpha\rangle_2$, $|\alpha\rangle_3$ by
injection of classical microwave fields. Subsequently a Rydberg
atom is initially prepared in the superposition
$(|e\rangle+|g\rangle)/\sqrt{2}$ by laser excitation to state
$|e\rangle$ followed by a $\pi/2$ pulse of classical microwave
radiation in the first Ramsey zone $R_1$. The $\pi/2$ pulses of
the Ramsey zones effect the transformations
$|e\rangle\rightarrow|e\rangle+|g\rangle$,
$|g\rangle\rightarrow|g\rangle-|e\rangle$. Thus as the atom enters
the cavity, the initial atom-field state is
\begin{equation}
|\psi(0)\rangle=\frac{1}{\sqrt{2}}(|e\rangle+|g\rangle)
|\alpha\rangle_1|\alpha\rangle_2|\alpha\rangle_3.
\end{equation}
Just as the atom leaves the cavity $C_1$, the state of the system
is
\begin{equation}
|\psi(t_1)\rangle=\frac{1}{\sqrt{2}}[|e\rangle|\alpha
e^{-i\theta_1}\rangle_1+|g\rangle|\alpha
e^{i\theta_1}\rangle_1]|\alpha\rangle_2|\alpha\rangle_3,
\end{equation}
where $\theta_1=\lambda t_1$, $t_1$ is the transit time of the
atom across the cavity $C_1$.
Just as the atom leaves the cavity $C_2$, the state of the system
is
\begin{eqnarray}
|\psi(t_1+t_2)\rangle=\frac{1}{\sqrt{2}}[|e\rangle|\alpha
e^{-i\theta_1}\rangle_1|\alpha e^{-i\theta_2}\rangle_2\nonumber\\
+|g\rangle|\alpha e^{i\theta_1}\rangle_1|\alpha
e^{i\theta_2}\rangle_2]|\alpha\rangle_3,
\end{eqnarray}
where $\theta_2=\lambda t_2$, $t_2$ is the transit time of the
atom across the cavity $C_2$. After the atom leaves the cavity
$C_3$, the state of the system
\begin{eqnarray}
|\psi(t_1+t_2+t_3)\rangle=&\frac{1}{\sqrt{2}}[|e\rangle|\alpha
e^{-i\theta_1}\rangle_1|\alpha e^{-i\theta_2}\rangle_2|\alpha
e^{-i\theta_3}\rangle_3\nonumber\\
&+|g\rangle|\alpha e^{i\theta_1}\rangle_1|\alpha
e^{i\theta_2}\rangle_2|\alpha e^{i\theta_3}\rangle_3],
\end{eqnarray}
where $\theta_3=\lambda t_3$, $t_3$ is the transit time of the
atom across the cavity $C_3$. Since the three cavities are
identical, it is reasonable to assume that $t_1=t_2=t_3=t$, and
therefore $\theta_1=\theta_2=\theta_3=\theta$. The atom is
velocity selected such that $\theta=\pi/2$.
\begin{figure}[hbtp!]
\centerline{\includegraphics[scale=0.8,angle=0]{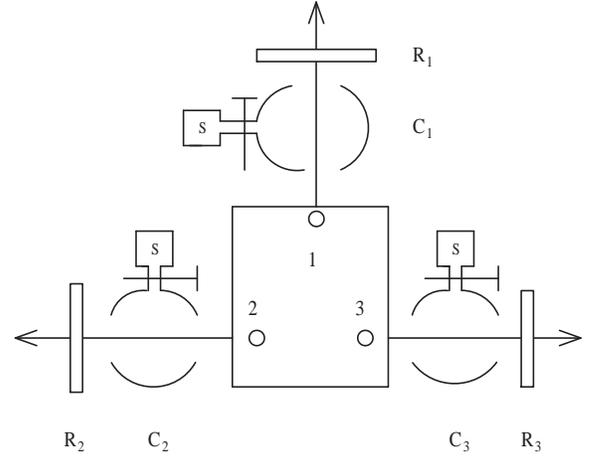}}
\caption{Experimental setup for generation of three-cavity
\emph{W}-type entangled coherent state. $C_1$, $C_2$, and $C_3$
are identical cavities. $R_1$, $R_2$ and $R_3$ are the Ramsey
zones. Atoms 1, 2, 3 are initially in $\emph{W}$ state
$|\emph{W}\rangle=\frac{1}{\sqrt{3}}(|egg\rangle+|geg\rangle+|gge\rangle)$.
$S$ is a classical source of microwaves which are injected into
the cavity through a waveguide. }\label{fig2}
\end{figure}
After the atom leaves the cavity $C_3$, the Ramsey zone $R_2$
applies a $\pi/2$ pulse to give
\begin{eqnarray}
|\psi^{'}(3t)\rangle&=N_0\{|g\rangle[|\beta\rangle_1|\beta\rangle_2|\beta\rangle_3
+|-\beta\rangle_1|-\beta\rangle_2|-\beta\rangle_3]\nonumber\\
&-|e\rangle[|\beta\rangle_1|\beta\rangle_2|\beta\rangle_3
-|-\beta\rangle_1|-\beta\rangle_2|-\beta\rangle_3]\},\nonumber\\
\end{eqnarray}
where $\beta=\alpha e^{-i\theta}$, and $N_0$ is the normalization
factor. If the atom is subsequently selectively ionized and found
to be in the ground state $|g\rangle$ or excited state
$|e\rangle$, then the three-cavity field is respectively projected
onto the state
\begin{equation}
|\beta\rangle_1|\beta\rangle_2|\beta\rangle_3+|-\beta\rangle_1|-\beta\rangle_2|-\beta\rangle_3,
\end{equation}
or the state
\begin{equation}
|\beta\rangle_1|\beta\rangle_2|\beta\rangle_3
-|-\beta\rangle_1|-\beta\rangle_2|-\beta\rangle_3,
\end{equation}
which are just the so-called GHZ-type ECSs.

Next, let us discuss how to prepare three-cavity \emph{W}-type
ECS. The experimental setup is shown in Fig.~\ref{fig2}. Assuming
three identical cavities $C_1$, $C_2$ and $C_3$ are initially in
coherent states $|\alpha_1\rangle$, $|\alpha_2\rangle$ and
$|\alpha_3\rangle$ by injection of classical microwave fields,
respectively. Atoms 1, 2, and 3 share an \emph{W} state, i.e.
$|\emph{W}\rangle_{123}=\frac{1}{\sqrt{3}}(|egg\rangle+|geg\rangle+|gge\rangle)$.
Firstly, we send atoms 1, 2 and 3 at the same velocity, which will
go through cavities $C_1$, $C_2$ and $C_3$, respectively. The
initial state of the three atoms and the three-cavity fields is
\begin{eqnarray}
|\Psi(0)\rangle=\frac{1}{\sqrt{3}}[|egg\rangle+|geg\rangle
+|gge\rangle]|\alpha\rangle_1|\alpha\rangle_2|\alpha\rangle_3.
\end{eqnarray}
After the three atoms leave their respective cavities, the state
of the whole system is
\begin{eqnarray}
|\Psi(\tau)\rangle=\frac{1}{\sqrt{3}}[|egg\rangle|\alpha
e^{-i\theta}\rangle_1|\alpha e^{i\theta}\rangle_2|\alpha
e^{i\theta}\rangle_3\nonumber\\
+|geg\rangle|\alpha e^{i\theta}\rangle_1|\alpha
e^{-i\theta}\rangle_2|\alpha e^{i\theta}\rangle_3\nonumber\\
+|gge\rangle|\alpha e^{i\theta}\rangle_1|\alpha
e^{i\theta}\rangle_2|\alpha e^{-i\theta}\rangle_3],
\end{eqnarray}
where $\theta=\lambda\tau$, $\tau$ is the transit time of the
three atoms across the cavities, respectively. (We assume the
transit times of the three atom are equal.) After the three atoms
1, 2, and 3 leave their respective cavities, they enter three
Ramsey zones $R_1$, $R_2$, $R_3$, respectively. The three Ramsey
zones apply $\pi/2$ pulses to give
\begin{eqnarray}
\label{e18}
|\Psi^{'}(\tau)\rangle=N_e[(|egg\rangle_{123}-|eeg\rangle_{123}
+|ggg\rangle_{123}-|geg\rangle_{123}\nonumber\\
-|ege\rangle_{123}+|eee\rangle_{123}-|gge\rangle_{123}\nonumber\\
+|gee\rangle_{123})|\alpha e^{-i\theta}\rangle_1|\alpha
e^{i\theta}\rangle_2|\alpha
e^{i\theta}\rangle_3\nonumber\\
+(|geg\rangle_{123}-|eeg\rangle_{123}+|ggg\rangle_{123}-|egg\rangle_{123}
-|gee\rangle_{123}\nonumber\\+|eee\rangle_{123}-|gge\rangle_{123}
+|ege\rangle_{123})|\alpha e^{i\theta}\rangle_1|\alpha
e^{-i\theta}\rangle_2|\alpha e^{i\theta}\rangle_3\nonumber\\
+(|ggg\rangle_{123}-|geg\rangle_{123}-|egg\rangle_{123}+|eeg\rangle_{123}
+|gge\rangle_{123}\nonumber\\+|eee\rangle_{123}-|ege\rangle_{123}
-|gee\rangle_{123})|\alpha e^{i\theta}\rangle_1|\alpha
e^{i\theta}\rangle_2|\alpha e^{-i\theta}\rangle_3],\nonumber\\
\end{eqnarray}
where $N_e$ is the normalization factor. We assume that the atom
is velocity selected such that $\theta=\pi/2$. Introducing
$\beta=\alpha e^{-i\theta}$, Eq.~(\ref{e18}) can be rewritten as
\begin{widetext}
\begin{eqnarray}
\label{e19} |\Psi^{'}(\tau)\rangle=N_e[(|ggg\rangle_{123}
+|eee\rangle_{123})(|-\beta\rangle_1|\beta\rangle_2|\beta\rangle_3
+|\beta\rangle_1|-\beta\rangle_2|\beta\rangle_3
+|\beta\rangle_1|\beta\rangle_2|-\beta\rangle_3)\nonumber\\
-(|gge\rangle_{123}+|eeg\rangle_{123})(|-\beta\rangle_1|\beta\rangle_2|\beta\rangle_3
+|\beta\rangle_1|-\beta\rangle_2|\beta\rangle_3
-|\beta\rangle_1|\beta\rangle_2|-\beta\rangle_3)\nonumber\\
+(|egg\rangle_{123}+|gee\rangle_{123})(|-\beta\rangle_1|\beta\rangle_2|\beta\rangle_3
-|\beta\rangle_1|-\beta\rangle_2|\beta\rangle_3
-|\beta\rangle_1|\beta\rangle_2|-\beta\rangle_3)\nonumber\\
-(|geg\rangle_{123}+|ege\rangle_{123})(|-\beta\rangle_1|\beta\rangle_2|\beta\rangle_3
-|\beta\rangle_1|-\beta\rangle_2|\beta\rangle_3
+|\beta\rangle_1|\beta\rangle_2|-\beta\rangle_3)].
\end{eqnarray}
\end{widetext}
 After the three atoms 1, 2 and 3 exit from their
respective cavities, their final states are analyzed in the
state-selective field-ionization detectors, respectively.
Depending on the detection results of the atomic states, the
three-cavity fields are projected onto one of the four
\emph{W}-type ECSs in Eq.~(\ref{e19}).

In order to realize the suggested schemes, it is necessary to
discuss dissipative process due to cavity losses and atomic
spontaneous emission. However, we consider Rydberg atoms of long
radiative lifetime and high-Q superconducting microwave cavities,
the atom-field interactions dominate the dissipative processes.
According to Ref.~\cite{Osnaghi}, for Rydberg atoms in circular
states with principal quantum numbers $50$ and $51$ (transition
frequency $\nu_0\sim51$ GHz), the atomic radiative lifetime
$T_{at}$ can reach $30$ ms and the cavity lifetime can be $T_r=1$
ms (corresponding to $Q=3\times10^{8}$). The length of the cavity
is of the order of centimeter and the velocity of the atom is of
the order of $100$ m/s, so the transit time of the atom in the
cavity is about $0.1$ ms.  This transit time is much shorter than
the atomic radiative lifetime $T_{at}$ and the cavity lifetime
$T_r$, therefore, the decoherence due to cavity losses and atomic
spontaneous emission can be ignored. Consequently, the GHZ-type
and \emph{W}-type ECSs of three-cavity fields can be generated.

In conclusion, we have proposed schemes for preparing the GHZ-type
and \emph{W}-type ECSs of three-cavity fields, and they can be
realized experimentally based on current cavity QED technique.
Furthermore, our schemes can be easily generalized to the
preparation of the GHZ-type and \emph{W}-type ECSs of $n$-cavity
fields ($n\geq3$).

This work was supported by the National Natural Science Foundation
of China (Grants No. 60178001 and No. 10074046)

Thanks for C. C. Gerry's comments about our manuscript.

\end{document}